# Performance Benefits of DataMPI: A Case Study with BigDataBench

Fan Liang[1,2] Chen Feng[1,2] Xiaoyi Lu[3] Zhiwei Xu[1]

[1]Institute of Computing Technology, Chinese Academy of Sciences
[2]University of Chinese Academy of Sciences, China
[3]Department of Computer Science and Engineering, The Ohio State University
{liangfan, fengchen, zxu}@ict.ac.cn, luxi@cse.ohio-state.edu

## Abstract

Apache Hadoop and Spark are gaining prominence in Big Data processing and analytics. Both of them are widely deployed on Internet companies. On the other hand, high-performance data analysis requirements are causing academical and industrial communities to adopt state-of-the-art technologies in HPC to solve Big Data problems. Recently, we have proposed a key-value pair based communication library, DataMPI, which is extending MPI to support Hadoop/Spark-like Big Data Computing jobs. In this paper, we use BigDataBench, a Big Data benchmark suite, to do comprehensive studies on performance and resource utilization characterizations of Hadoop, Spark and DataMPI. From our experiments, we observe that the job execution time of DataMPI has up to 55% and 39% speedups compared with those of Hadoop and Spark, respectively. Most of the benefits come from the high-efficiency communication mechanisms in DataMPI. We also notice that the resource (CPU, memory, disk and network I/O) utilizations of DataMPI are also more efficient than those of the other two frameworks.

## 1. Introduction

Data explosion is becoming an irresistible trend with the development of Internet, social network, e-commerce, etc. Over the last decade, there have been emerging a lot of systems and frameworks for Big Data, such as MapReduce [7], Hadoop [1], Dyrad [9], Yahoo! S4 [15] and so on. Apache Hadoop has become as the defacto standard for Big Data processing and analytics. Many clusters in the production environment already contain thousands of nodes to dedicatedly run Hadoop jobs everyday. Beyond the success of Hadoop on its scalability and fault-tolerance, Apache Spark [20] provides another feasible way to process large amount of data by introducing the in-memory computing techniques. Nowadays, both of them have attracted more and more attentions from academical and industrial areas.

However, the performance of current commonly used Big Data systems is still in a sub-optimal level. Many studies [10, 13, 17, 18] have been trying to adopt state-of-the-art technologies in the High Performance Computing (HPC) area to accelerate the performance of Big Data processing. As one example of these attempts, our previous work [12, 14, 19] shows the performance of Hadoop communication primitives still have huge performance improvement potentials, and Message Passing Interface (MPI), which is widely used in the field of HPC, can help to optimize communication performance of Hadoop. Furthermore, we have proposed a key-value pair based communication library, DataMPI [2, 14], to efficiently execute Hadoop/Spark-like Big Data Computing jobs by extending MPI. Since the open-source nature of these systems, it will be very interesting for the users to know the performance characteristics of Hadoop, Spark, and DataMPI by doing a systematical performance evaluation on different aspects.

But one of the most important issues is how to evaluate these various innovative Big Data systems in a fair and comprehensive way. Many benchmarks have been proposed to help this kind of evaluations for fitting different application scenarios, such as MR-Bench [11], HiBench [8], YCSB [6], BigDataBench [16] and so on. In this paper, we choose BigDataBench, a representative benchmark suite, to evaluate Hadoop, Spark, and DataMPI. It serves our three needs well: a) typical workloads among a wide range of application domains; b) providing tools to generate large volume data efficiently; c) preserving real-world data characteristics in the generated data.

Based on BigDataBench, our study shows through high performance communication mechanisms, DataMPI can achieve up to 55% and 39% improvements compared with Hadoop and Spark, respectively. By profiling resource utilization, we observe the DataMPI library can leverage system resources more efficiently than the other two frameworks.

The rest of this paper is organized as follows. Section 2 discusses background and related work. Section 3 states our experiments methodology. The evaluation results and analysis are given in Section 4. Section 5 concludes the paper.

## 2. Background and Related Work

### 2.1 Hadoop

Hadoop [1] is an open-source implementation of the MapReduce programming model, which is expressive to capture a wide class of computations, and has been widely used in various areas and applications, such as log analysis, machine learning, search engine, etc. The success of Hadoop owes to its high scalability, built-in fault-tolerance and simplicity of programming. A MapReduce job of Hadoop contains several Map/Reduce tasks. Map tasks emit key/value pairs to the corresponding Reduce tasks according to the partition function. Reduce tasks receive and sort the intermediate data, process the values of the same key, then output the results.

### 2.2 Spark

Spark [20] is a cluster computing system which can perform analytics with in-memory techniques. Different from traditional in-memory systems, Spark aims at task-parallel jobs, especially iterative algorithms in machine learning and interactive data mining. The benefits of Spark are mainly contributed from the distributed memory abstraction, resilient distributed datasets (RDDs), which support coarse-grained transformations. RDDs access the process heap directly without any system calls to access the memory, which makes them efficient. To support fault tolerance, an RDD logs the transformations to build a dataset as its lineage, which makes the lost RDD have enough information to recover from other RDDs.

### 2.3 DataMPI

DataMPI [14] is a key-value pair based communication library which extends MPI for Hadoop/Spark-like Big Data Computing systems. Different from the buffer-to-buffer communication in MPI, DataMPI adopts the key-value based communication which captures the essential communication nature in Hadoop/Spark-like Big Data Computing systems. DataMPI implements a bipartite communication model which supports four communication characteristics of Big Data Computing systems, dichotomic, dynamic, data-centric, and diversified. A job of DataMPI contains several tasks which are divided into O/A communicators and form a bipartite graph in the underlying communication. Data movement from O communicator to A communicator is scheduled implicitly by the library. Concurrent running tasks are dynamically scheduled to the corresponding communicators. For supporting data-centric, DataMPI partitions and stores the emitted data by O tasks in memory or disk. Then, A tasks are scheduled to the corresponding worker processes to read the intermediate data locally. Data movement is pipelining with the computation overlapped in O tasks. By buffering the intermediate data in worker processes, DataMPI can accelerate the execution performance without redundant disk I/O operations. DataMPI also supports fault tolerance by key-value pair based checkpoint/restart.

### 2.4 BigDataBench

BigDataBench [16] is a benchmark suite for Big Data Computing. It is abstracted from typical Internet service workloads and covers representative data sets and broad applications. BigDataBench summarizes six emblematical application scenarios, including three important application domains, e.g. search engine, social networks, and e-commerce, and three basic operation scenarios, e.g. micro-benchmark, relational query, and basic datastore operations "Cloud OLTP". These scenarios contain nineteen typical application/algorithm workloads, which consider online service, offline service and real-time analytics. The spectrum of data types of these workloads can be classified into structured, semi-structured, and unstructured data with text, table, or graph data formats. BigDataBench provides a data generator for benchmarks based on real life data sets, and generates the six seed models by extracting the synthetic characteristics of corresponding real-world data sets, including wikipedia entries, amazon movie reviews, e-commerce transaction data and so on. Users can generate synthetic data by scaling the seed models while keeping the characteristics of data.

### 2.5 Other Related Works

Recent studies [17, 18] show high performance interconnects, like InfiniBand [3], can be used to improve the performance of Apache Hadoop. Several technologies have been used in these works including in-memory merging, pipelining, pre-fetching and caching. Hadoop-RDMA [10, 13, 18] is one of the representative systems and achieves significant performance benefits through RDMA-capable interconnects with enhanced designs of various components inside Hadoop, such as HDFS, RPC, MapReduce.

## 3. Benchmarking Methodology

In this section, we present our workloads chosen from BigDataBench and our evaluation methodology.

### 3.1 Chosen Workload

We choose WordCount, Grep and Sort as the basic operation micro-benchmarks, Naive Bayes and K-means as application benchmarks.

| No. | Workload | Type |
|---|---|---|
| 1 | Sort | Micro-benchmark |
| 2 | WordCount | |
| 3 | Grep | |
| 4 | Naive Bayes | Social Network |
| 5 | K-means | E-commerce |

**Table 1.** Representative Workloads

- **Basic Operations**: WordCount, Grep and Sort are fundamental and widely used operations in broad analysis processes. *WordCount* counts the number of each word occurrences in a collection of documents. *Grep* searches strings conforming to a certain pattern in the input documents and counts the number of the occurrence of the matched strings. *Sort* sorts the records of input files based on the value of keys. We use two input data sets while running Sort benchmark. One is Normal Sort with compressed sequence input data, the other is Text Sort with uncompressed text input data.
- **Applications**: K-means and Naive Bayes are typical applications in social network and e-commerce scenarios. *K-means* is a classical clustering algorithm in data mining which aims to partition the input objects to $k$ clusters by calculating the nearest mean cluster of each object belongs to. The algorithm firstly chooses the initial cluster centers, then uses iterative refinement techniques to update the centroids until the bias of each centroid between two successive iterations is less than a given threshold value. Each iteration procedure contains two steps: assignment and update. *Naive Bayes* is a probabilistic algorithm for classification. It is based on Bayes' theorem with strong independence assumptions, which means the features of the model are independent with each other. The basic computation of Naive Bayes includes calculating term frequency and maximum likehood. This algorithm contains two steps: model training and classification.

  Both of the two applications have various computation characteristics, which can be used to evaluate the comprehensive performance of a Big Data system.

### 3.2 Evaluation Methodology

We follow a seven-pronged approach to evaluate the performance of Hadoop, Spark, and DataMPI as shown in Figure 1. We start the performance comparison using three micro-benchmarks. Then, we analyze the resource utilization of the systems with two micro-benchmarks from four aspects including CPU utilization, network I/O throughput, disk I/O throughput and memory footprint, and calculate the efficiency of CPU and memory. To measure the effects of system overhead, we evaluate the performance when processing small jobs. Finally, we use two application benchmarks to evaluate the comprehensive performance. We summarize the results along these seven dimensions in Section 4.7.

## 4. Experimental Evaluation

### 4.1 Experiment Setup

We use a cluster composed of 8 nodes interconnected by a 1 Gigabit Ethernet switch as our testbed. Each node is equipped with two Intel Xeon E5620 CPU processors (2.4 GHz) with enabling the hyper-threads. Each core has private L1 and L2 caches, and all cores share the L3 cache. Each node has 16 GB DDR3 RAM with 1333 MHz and one 150 GB free space SATA disk. Table 2 shows the detailed hardware configurations.

The operation system used is CentOS release 6.5 (Final) with kernel version 2.6.32-431.el6.x86_64. The softwares used are

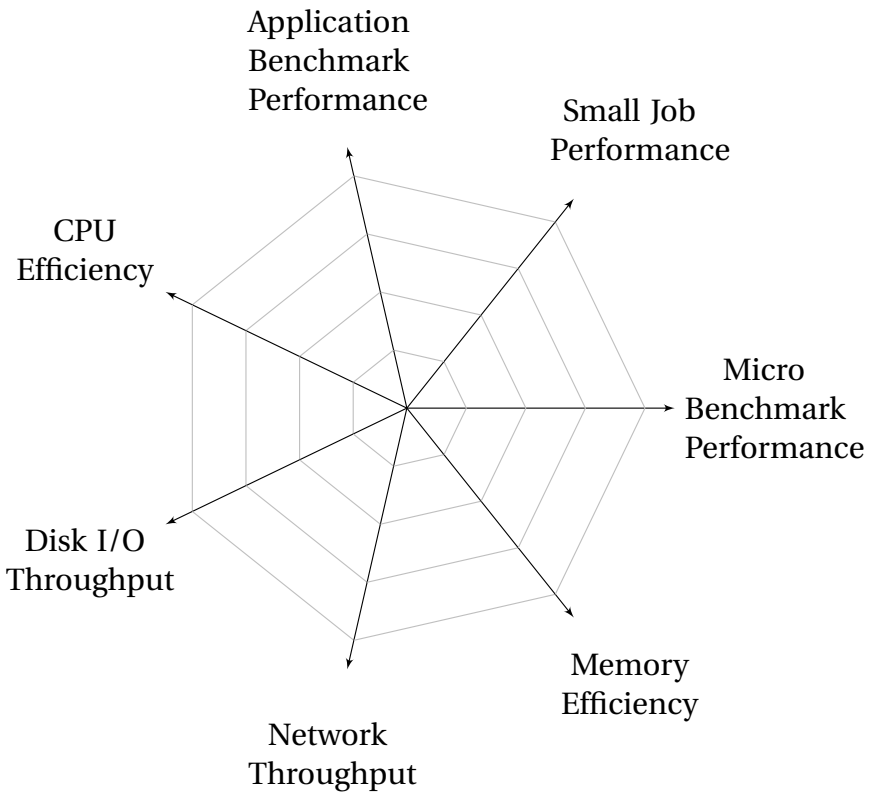

**Figure 1.** Evaluation Methodology

| | |
|---|---|
| CPU type | Intel ® Xeon E5620 |
| # cores | 4 cores @2.4G |
| # threads | 8 threads |
| # sockets | 2 |
| L1 I/D Cache | 32 KB |
| L2 Cache | 256 KB |
| L3 Cache | 12 MB |
| Memory | 16 GB |
| Disk | 150GB free SATA disk |

**Table 2.** Details of Hardware Configuration

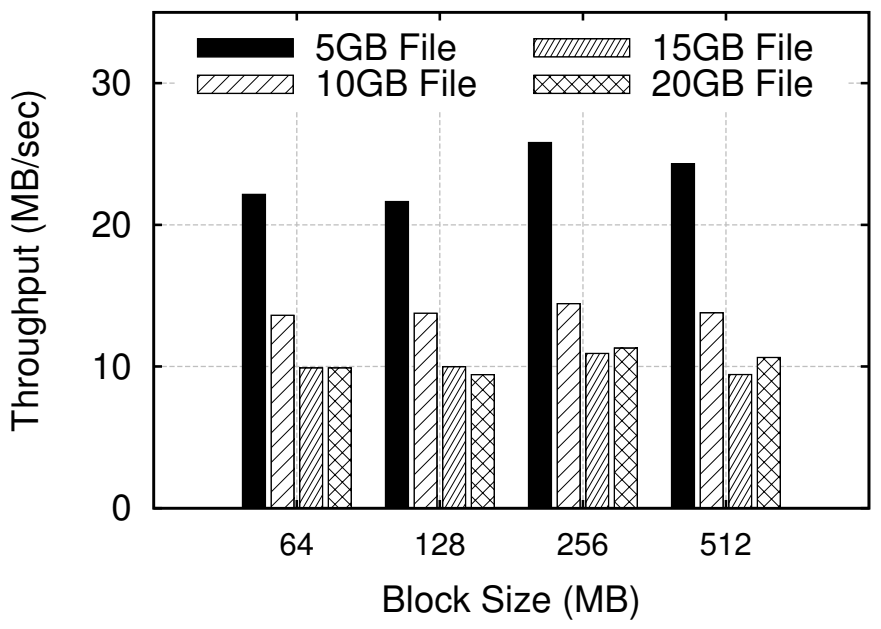

(a) HDFS Block Size Tuning based on DFSIO

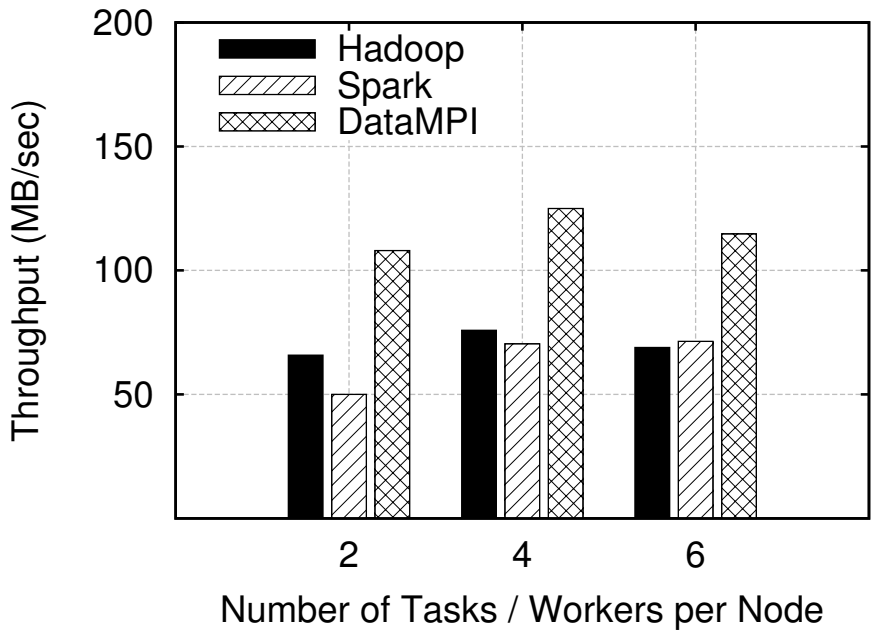

(b) Number of Tasks / Works Tuning based on Text Sort

**Figure 2.** Parameter Tuning

JDK 1.7.0_25, Hadoop 1.2.1, Mahout 0.8 [4], Spark 0.8.1, Scala 2.9.3 and BigDataBench 2.1. The MPI implementation is MVAPICH2-2.0b. For all evaluations, we report results that are average across three executions.

### 4.2 Chosen Parameters

Hadoop, Spark and DataMPI have abundant parameters to set in different systems and clusters. In this section, we tune the parameters for fair evaluations. Among most parameters, we mainly focus on the HDFS block size and the number of tasks / workers, because the disk and network will easily become the bottleneck in our testbed.

We use DFSIO program, a file system level benchmark of Hadoop, as the workload for tuning HDFS block size, and Text Sort benchmark for tuning the number of concurrent tasks / workers. We vary the HDFS block size from 64 MB to 512 MB with input data size from 5 GB to 20 GB, and measure the throughput. Figure 2(a) shows when block size is 256 MB, the throughput achieves the best. When tuning the number of concurrent tasks / workers, we measure the Text Sort throughput by processing 1 GB data per Hadoop/DataMPI task and 128 MB data per Spark worker with increasing the number of concurrent tasks / workers from 2 to 6 per node. With this configuration, we can execute the Spark tests without OutOfMemory Errors. Figure 2(b) shows these systems can get the best throughput when the number of tasks / workers on each node is 4.

Based on the two tests, we run our following evaluation based on 256 MB HDFS block size with 3 replications and 4 concurrent tasks / workers per node.

### 4.3 Micro-benchmark Performance

In this section, we evaluate the performance of micro-benchmarks among Hadoop, Spark, and DataMPI. We use BigDataBench Text Generator to produce the text data set. The seed model used in Text Generator is *lda_wiki1w* trained from wikipedia entries corpus.

The text data set is used for Text Sort, WordCount and Grep. The input of Normal Sort is sequence data, which is converted from text data by *ToSeqFile* of BigDataBench. *ToSeqFile* runs a MapReduce job and copies each line of the input data to the key and value, then compresses the output with GzipCodec. We vary the input data size from 4 GB to 64 GB.

We evaluate Spark with Text Sort, WordCount and Grep workloads. Spark fails for OutOfMemory Errors when testing Normal Sort and Text Sort except the 8 GB case of Text Sort, even though we allocate the memory to each worker as large as possible. As shown in Figure 3(a) and Figure 3(b), DataMPI has 29%-33% improvement in Normal Sort, and 34%-42% improvement in Text Sort compared to Hadoop. In the case of 8 GB Text Sort, DataMPI costs 69 seconds, which is 39% faster than 114 seconds in Spark. Figure 3(c) shows the result of WordCount. DataMPI and Spark have similar performance. Both of them have 47%-55% improvements than Hadoop. A detailed analysis of Sort and WordCount evaluations from the view of resource utilization is given in Section 4.4. The Grep evaluation result shown in Figure 3(d) exhibits that DataMPI cuts down the execution time by 33%-42% compared to Hadoop and 19%-29% compared to Spark.

According to the results of micro-benchmark evaluations, DataMPI has averagely 40% improvement than Hadoop and 14% improvement than Spark.

### 4.4 Profile of Resource Utilization

In this section, we measure the resource utilization of Hadoop, Spark, and DataMPI. The workloads of 8 GB Text Sort and 32 GB WordCount are chosen to be profiled from four aspects, CPU utilization, disk throughput, network throughput, and memory footprint. We record the total usage percentage and the CPU wait I/O

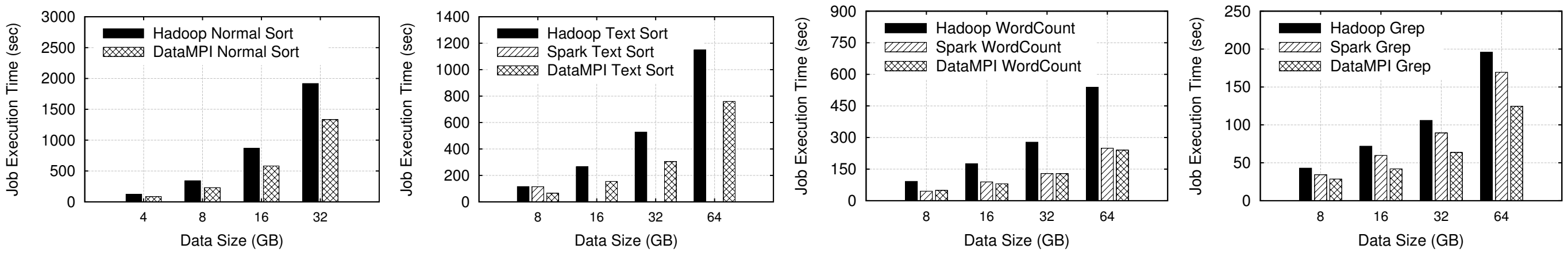

(a) Normal Sort (b) Text Sort (c) WordCount (d) Grep

**Figure 3.** Performance Comparison of Different Micro-benchmarks

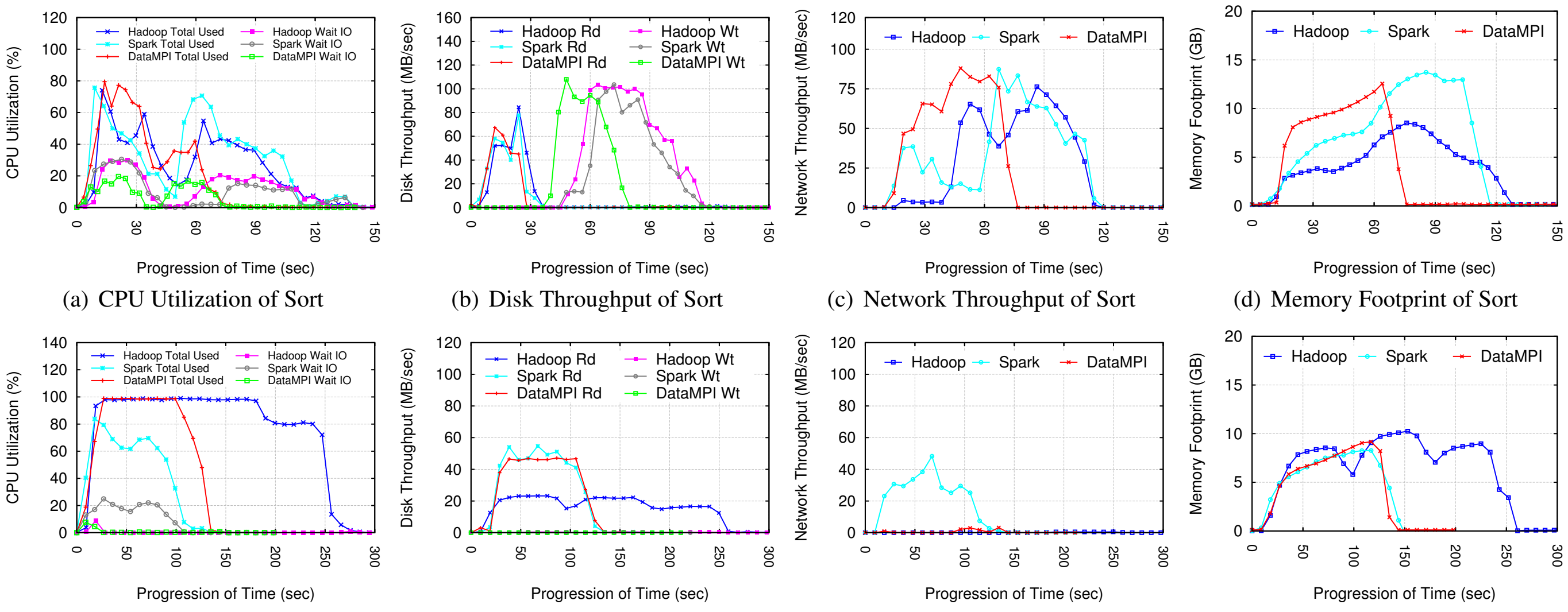

(a) CPU Utilization of Sort (b) Disk Throughput of Sort (c) Network Throughput of Sort (d) Memory Footprint of Sort

(e) CPU Utilization of WordCount (f) Disk Throughput of WordCount (g) Network Throughput of WordCount (h) Memory Footprint of WordCount

**Figure 4.** Resource Utilization of Different Benchmarks

percentage. A high CPU wait I/O percentage means CPU costs most of time to wait for I/O operations to complete.

In the 8 GB Text Sort case, DataMPI costs 69 seconds while Hadoop and Spark cost 117 seconds and 114 seconds, respectively. The O phase of DataMPI costs 28 seconds (O tasks execute during O phase), the Map phase of Hadoop costs 36 seconds and the Stage 0 of Spark costs 38 seconds (During Stage 0, Spark loads the data from HDFS and creates the RDD for next execution). The result of CPU utilization is shown in Figure 4(a). Considering the average CPU utilization during 0-117 secondes, DataMPI, Spark, and Hadoop get 24%, 38%, and 37%. The average CPU wait I/O percentages of DataMPI, Spark, and Hadoop are 6%, 12%, and 15%. Figure 4(b) shows the disk throughput. The average disk read throughputs of DataMPI O phase, Hadoop Map phase, and Spark Stage 0 are 50 MB/sec, 49 MB/sec, and 46 MB/sec. The average disk write throughputs of DataMPI, Hadoop, and Spark are 69 MB/sec, 67 MB/sec, and 66 MB/sec. This means these systems have similar disk I/O throughput performance in this test case. Figure 4(c) shows the network throughput. DataMPI achieves averagely 62 MB/sec, which is 59% higher than 39 MB/sec in Hadoop and 55% higher than 40 MB/sec in Spark. Besides the HDFS communication, we should notice that the communication caused by data movement from O communicator to A communicator mainly happens in DataMPI O phase. Figure 4(d) shows the memory footprint. The average memory usages of DataMPI, Spark, and Hadoop during 0-117 seconds are 5 GB, 9 GB, and 5 GB.

In the case of WordCount, DataMPI and Spark cost almost the same execution time, 130 seconds, and improve the total execution time by 53% compared to 275 seconds in Hadoop. Figure 4(e) shows the CPU utilization. Considering the average CPU utilization during 0-275 seconds, DataMPI, Spark, and Hadoop get 47%, 30%, and 80%. Spark also has 8% CPU wait I/O. Figure 4(f) shows the average read throughputs of DataMPI and Spark are nearly 44 MB/sec which is much more efficient than 20 MB/sec in Hadoop. Figure 4(g) shows both of DataMPI and Hadoop have few network overhead while Spark has averagely 25 MB/sec throughput. There are two reasons. One is that the O/Map tasks read the HDFS data locally and do not have network communication which is different from Spark. The other is that the word dictionary of the input files is small and few intermediate data is generated. Figure 4(h) shows the average memory usages of DataMPI, Spark, and Hadoop during 0-275 seconds are 5 GB, 5 GB, and 9 GB.

From these cases, we observe that DataMPI can achieve higher network throughput than Hadoop and Spark. Compared to Hadoop, both of DataMPI and Spark can use memory more efficiently to cache intermediate data to reduce execution time.

### 4.5 Small Jobs

According to [5], more than 90% of MapReduce jobs in Facebook and Yahoo! are small jobs. The input data sizes of these jobs are usually kilo or mega bytes. The system overhead of initialization and finalization has a serious impact on performance of these jobs.

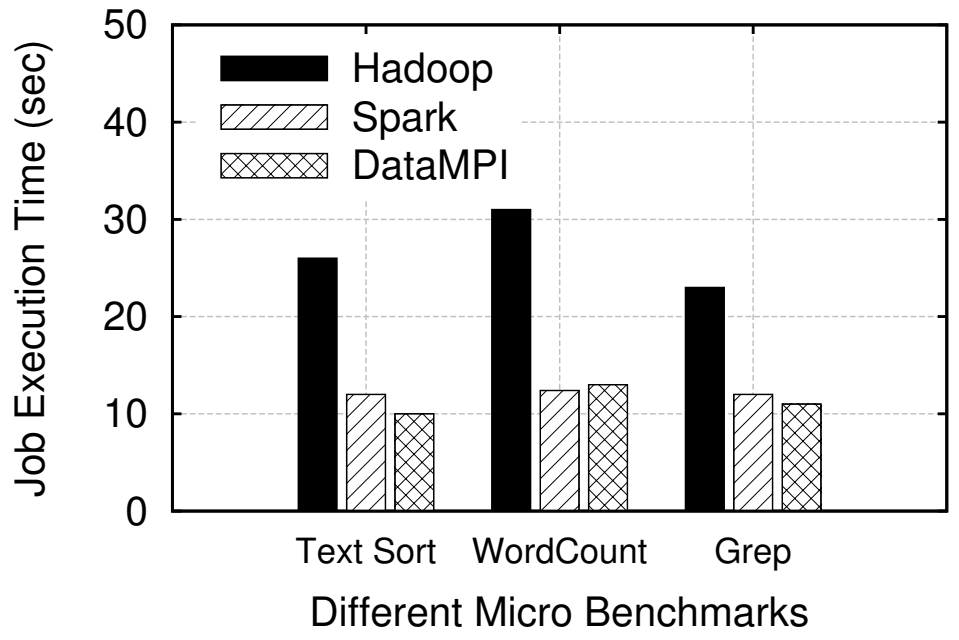

**Figure 5.** Performance Comparison Based on Small Jobs

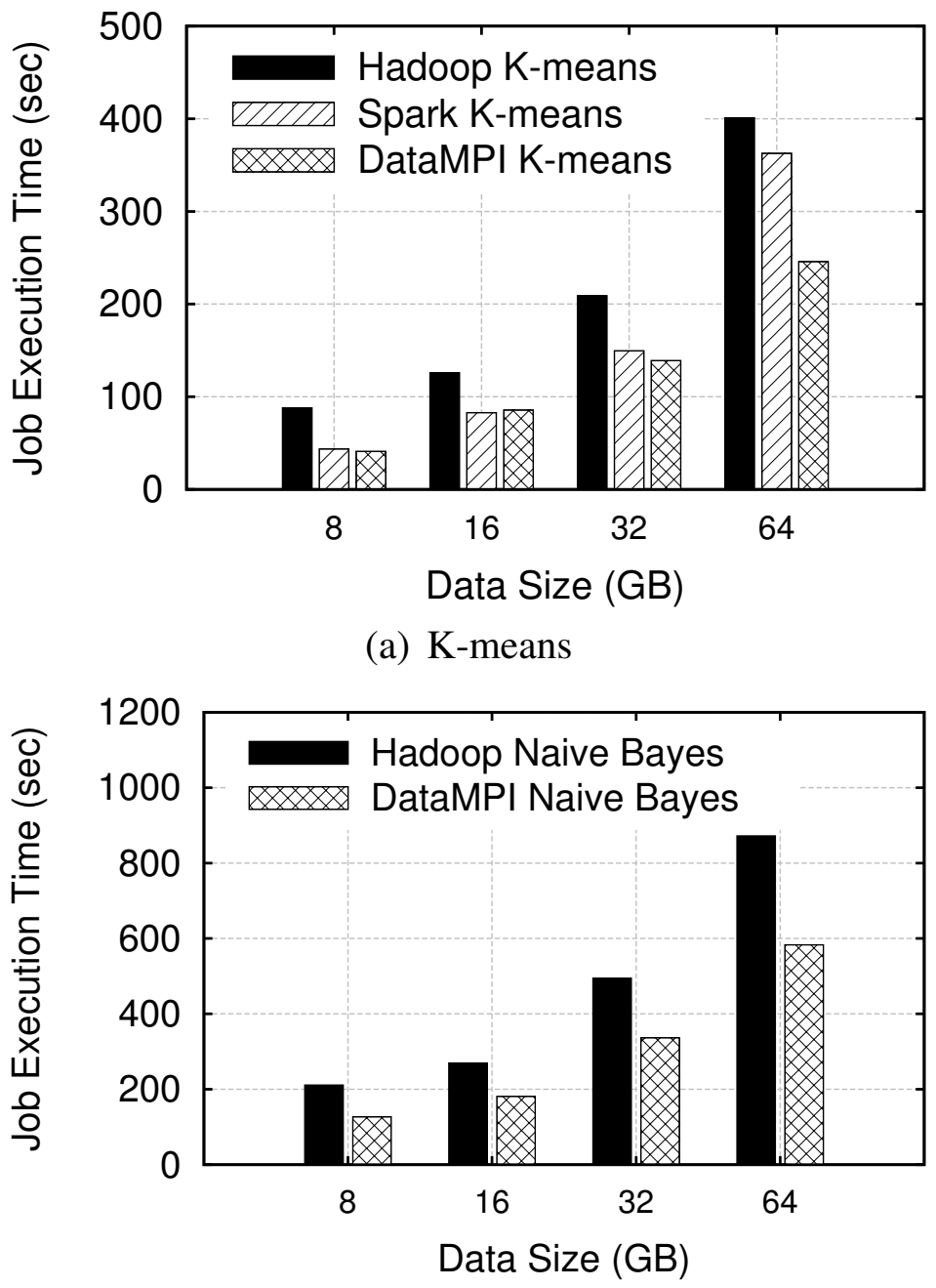

(a) K-means

(b) Naive Bayes

**Figure 6.** Performance Comparison of Application Benchmarks

In this section, we compare the performance of Hadoop, Spark, and DataMPI when doing the micro-benchmarks of Text Sort, WordCount and Grep. The input data of each benchmark is 128 MB. The number of the concurrent tasks/works is one per node. Figure 5 shows that DataMPI has similar performance with Spark, and is averagely 54% more efficient than Hadoop.

### 4.6 Application Benchmark Performance

In this section, we present the results of the application evaluations. The implementations of K-means and Naive Bayes for Hadoop used by BigDataBench come from Mahout. We first explain the processing characteristics of these applications from the implementation-level and then give the performance results.

**K-means:** The input data set is generated by Text Generator, in which five seed models, amazon1-amazon5, are used. Using *genData_Kmeans* of BigDataBench, text files are converted to sequence files from directory, then to the sparse vectors which are the input data of training clusters. Our evaluations are based on the sparse vectors and mainly focus on the performance of training execution. As stated in Section 3.1, K-means trains the cluster centroids iteratively. Each iterative execution in Mahout is a MapReduce job. In one job, Map tasks read the initial or previous cluster centroids from HDFS, afterwards, assign the input vectors to appropriate clusters according to the distance calculation and train the new centroids independently. At the end of Map tasks, new centroids will be sent to the Reduce tasks according to the cluster indexes. Reduce tasks receive and update the centroids for next iteration. We observe that most of K-means calculation happens in Map phase, and few intermediate data is generated. We transplant the Mahout K-means actuating logic to DataMPI and integrate some Mahout basic data structures in the implementation of DataMPI K-means.

Our tests show Spark have outstanding performance when doing the iteration computations after caching the data in the RDDs. For fair comparison with Hadoop, we record the execution time of the first iteration from the job start, which considering the overhead of loading data, computing and outputing results. Figure 6(a) shows that DataMPI has at most 39% improvement than Hadoop and 33% improvement than Spark when the input data size varies from 8 GB to 64 GB. In the future, we will give a detail performance comparison between Spark and DataMPI in the iterative applications.

**Naive Bayes:** We also use Text Generator to generate the input document set. By default, these documents are classified into five categories according to their dependent seed models, e.g. amazon1-amazon5. The procedure of Naive Bayes mainly contains two steps, including converting sequence files to sparse vectors and training the Naive Bayes model. Mahout runs several MapReduce jobs to create the sparse vectors. Firstly, one document is converted to a token array. After that, some MapReduce jobs are launched to count the term frequency in one document and document frequency of all terms. The sparse vector of one document is calculated according to the term frequency and document frequency. The main operation in steps above is counting, including term counting and document counting, which means that the characteristics of Naive Bayes is similar to WordCount. In our evaluation cases, the data sizes of sparse vector and term-counting dictionary are within several mega bytes. The model training contains two MapReduce jobs which do the probabilistic computing. These two jobs cost less time than the sparse vectors creating for the simple calculating and small input data size.

The latest BigDataBench lacks the implementation of Naive Bayes in Spark. We only compare the performance of this benchmark between DataMPI and Hadoop. Figure 6(b) shows DataMPI has 33 % improvement than Hadoop averagely, which is contributed by the high efficient data communication and computation of DataMPI in term occurrence counting and term frequency calculation, as we have explained in Section 4.4.

### 4.7 Discussion of Performance Results

We summarize the performance comparisons with different benchmarks using seven-pronged diagram presented earlier, depicted in Figure 7. Compared to Hadoop, DataMPI can averagely achieve 40%, 54%, and 36% performance improvements when running micro-benchmarks, small jobs, and application benchmarks. Compared to Spark, DataMPI can achieve 14% and 33% performance improvements when running micro-benchmarks and application benchmarks. DataMPI and Spark have similar performance when running small jobs. From the two cases, the average CPU utilizations of DataMPI, Spark, and Hadoop are 35%, 34%, and 59%, which means DataMPI and Spark can use CPU 39% and 41% more efficiently than Hadoop. DataMPI and Spark achieve similar disk I/O throughput which have averagely 49% improvement compared to Hadoop. DataMPI achieves 55% and 59% network throughput improvements than Spark and Hadoop, respectively. We observe both DataMPI and Spark can efficiently utilize memory to accel-erate the execution performance. These benefits of DataMPI come from the high performance communication design which is able to leverage system resources to pipeline the computation and communication operations efficiently.

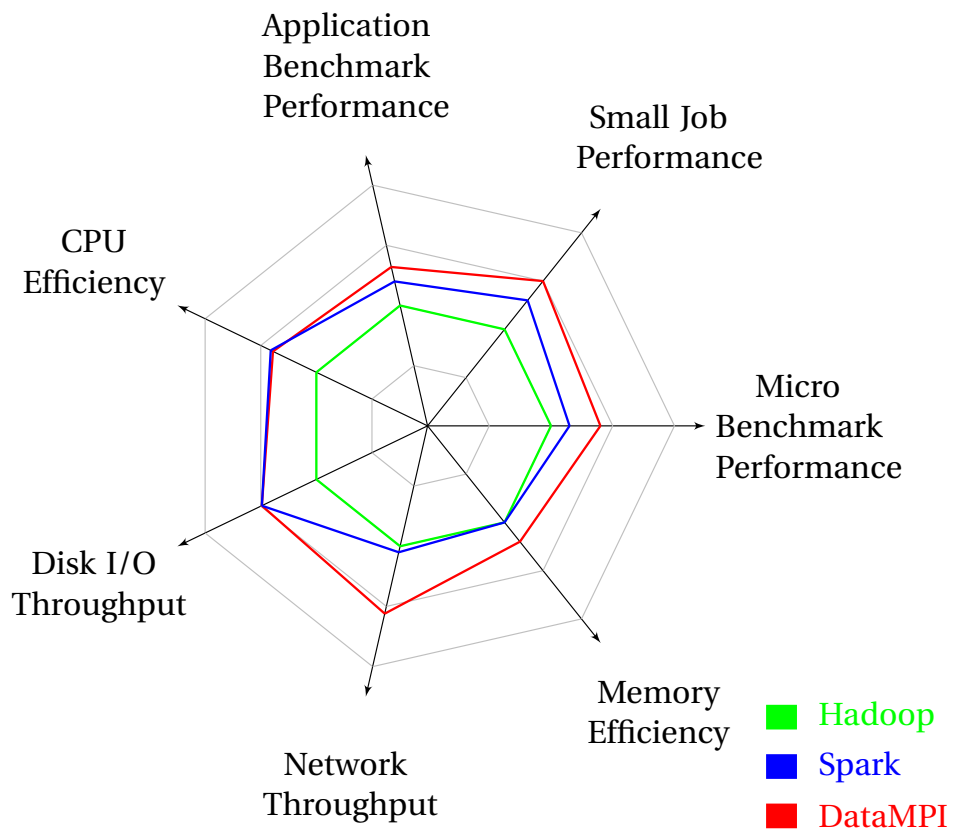

**Figure 7.** Evaluation Results

## 5. Conclusion

In this paper, we provide a comprehensive performance evaluation of Hadoop, Spark, and DataMPI based on BigDataBench. We choose three micro benchmarks (Sort, WordCount and Grep) and two application benchmarks (K-means and Naive Bayes) as our workloads. Based on two typical micro-benchmark cases, we present a detailed resource utilization analysis of the three systems. Our evaluation shows DataMPI can achieve 29 %-55 % performance improvements compared to Hadoop with the mircobenchmarks, and up to 39 % performance improvements compared to Spark. The small job tests show the overheads of DataMPI and Spark are low, and this makes them gain 54 % performance improvement compared to Hadoop. Evaluations of Naive Bayes and K-means benchmarks show DataMPI can achieve 33%-39% application-level performance than Hadoop and Spark.

## 6. Acknowledgments

We are very grateful to Dr. Lei Wang and Zijian Ming for their help to support this research, and also to the anonymous reviewers. This work is supported in part by the Strategic Priority Program of Chinese Academy of Sciences (Grant No. XDA06010401).